\documentclass[a4paper,12pt]{article}
\usepackage{amsmath}
\usepackage{latexsym}
\newcommand{\be}{\begin{eqnarray}}
\newcommand{\ee}{\end{eqnarray}}
\newcommand{\bdm}{\begin{displaymath}}
\newcommand{\edm}{\end{displaymath}}
\begin{document}
\title{\Large\textbf{Conditional Symmetries and the
Quantization of Bianchi Type I Vacuum Cosmologies with and without
Cosmological Constant}}
\author{\textbf{T. Christodoulakis}\thanks{e-mail:
tchris@cc.uoa.gr}~~~\textbf{T. Gakis}~~\textbf{\& ~G. O.
Papadopoulos}\thanks{e-mail: gpapado@cc.uoa.gr}}
\date{}
\maketitle
\begin{center}
\textit{University of Athens, Physics Department\\
Nuclear \& Particle Physics Section\\
Panepistimioupolis, Ilisia GR 157--71, Athens, Hellas}
\end{center}
\vspace{1cm} \numberwithin{equation}{section}
\begin{abstract}
In this work, the quantization of the most general Bianchi Type I
geometry, with and without a cosmological constant, is considered.
In the spirit of identifying and subsequently removing as many
gauge degrees of freedom as possible, a reduction of the initial
6--dimensional configuration space is presented. This reduction is
achieved by imposing as additional conditions on the wave
function, the quantum version of the --linear in momenta--
classical integrals of motion (conditional symmetries). The vector
fields inferred from these integrals induce, through their
integral curves, motions in the configuration space which can be
identified to the action of the automorphism group of Type I, i.e.
$GL(3,\Re)$. Thus, a wave function depending on one degree of
freedom, namely the determinant of the scale factor matrix, is
found.\\
A measure for constructing the Hilbert space is proposed. This
measure respects the above mentioned symmetries, and is also
invariant under the classical property of covariance under
arbitrary scalings of the Hamiltonian (quadratic constraint).
\end{abstract}
\newpage
\section{Introduction}
Since the conception by Einstein of General Relativity Theory, a
great many efforts have been devoted by many scientists to the
construction of a consistent quantum theory of gravity. These
efforts can de divided into two main approaches:
\begin{itemize}
\item[(a)] perturbative, in which one splits the metric into a background
(kinematical) part and a dynamical one:
$g_{\mu\nu}=\eta_{\mu\nu}+h_{\mu\nu}$ and tries to quantize
$h_{\mu\nu}$. The only conclusive results existing, are that the
theory thus obtained is highly nonrenormalizable \cite{goroff}.
\item[(b)] non perturbative, in which one tries to keep the twofold role
of the metric (kinematical and dynamical) intact. A hallmark in
this direction is canonical quantization.
\end{itemize}
In trying to implement this scheme for gravity, one faces the
problem of quantizing a constrained system. The main steps one has
to follow are:
\begin{itemize}
\item[(i)] define the basic operators $\widehat{g}_{\mu\nu}$
and $\widehat{\pi}^{\mu\nu}$ and the canonical commutation
relation they satisfy.
\item[(ii)] define quantum operators $\widehat{H}_{\mu}$ whose
classical counterparts are the constraint functions $H_{\mu}$.
\item[(iii)] define the quantum states $\Psi[g]$ as the common null
eigenvector of $\widehat{H}_{\mu}$, i.e. those satisfying
$\widehat{H}_{\mu}\Psi[g]=0$. (As a consequence, one has to check
that $\widehat{H}_{\mu}$, form a closed algebra under the basic
Canonical Commutation Relations (CCR).)
\item[(iv)] find the states and define the inner product in the space of
these states.
\end{itemize}
It is fair to say that the full program has not yet been carried
out, although partial steps have been made \cite{zanelli}.
Concerning point (iii) we deem it pertinent to clarify the meaning
of the imposition of the quantum constraints upon $\Psi[g]$. A
straightforward (modulo regularization prescriptions) but tedious
calculation shows that any functional which is not a scalar
functional of the curvature invariants (see \cite{munoz}) does not
solve the linear constraints. Therefore, the imposition of the
linear constraints, ensures that the wave functional will be a
(scalar) functional of the 3-geometry and not of the coordinate
system. Then, the dynamical evolution is provided by the quadratic
constraint; the consistency of the quantum algebra, guarantees
that the final wave functional, will be independent of the 4
dimensional coordinate system.

In the absence of a full solution to the problem, people have
turned to what is generally known as quantum cosmology. This is an
approximation to quantum gravity in which one freezes out all but
a finite number of degrees of freedom, and quantizes the rest. In
this way one is left with a much more manageable problem that is
essentially quantum mechanics with constraints. Over the years,
many models have appeared in the literature \cite{halliwell}. In
most of them, the minisuperspace is flat and the gravitational
field is represented by no more than three degrees of freedom
(generically the three scale factors of some anisotropic Bianchi
Type model \cite{amsterdamski}).

In a series of earlier publications \cite{Chris}, we have
considered the quantization of the most general spatially
homogeneous spacetime for various Bianchi Types. Thus, our --in
principle-- dynamical variables were the scale factors
$\gamma_{\alpha\beta}(t)$, the lapse function $N(t)$ and the shift
vector $N^{\alpha}(t)$.\\
The presence of the linear constraints --along with the
conditional symmetries of the corresponding Hamiltonian-- enabled
a reduction of the initial configuration space to a lower
dimensional one, spanned by the curvature invariants
characterizing the 3-geometry. The ultimate justification of this
reduction is the fact that --from the point of view of the
3-geometry-- the omitted degrees of freedom, are not physical but
gauge \cite{diff}.

The case of Bianchi Type I geometries, has been repeatedly treated
in the literature --both at the classical level \cite{Landau} and
the quantum level \cite{Ashtekar}. In all these works, the scale
factor matrix is taken to be diagonal. The r\^{o}le of the
topology of the 3-slices, is emphasized in the work of Ashtekar
and particularly in that of Hervik where a non trivial $T^{3}$
topology leads to a nine dimensional moduli space. The main reason
for this plethora of treatments of Type I, is the simplicity
brought by the vanishing structure constants, i.e. the high
spatial symmetry of the model. It is true that at the classical
level, the scale factor matrix, can be diagonalized on mass-shell
--through a constant matrix \cite{timeaut}-- while the shift can
be set equal to zero. However, if one intends to give weight to
all states, one has to start with the most general form which is
described by the 6 scale factors $\gamma_{\alpha\beta}(t)$ and the
lapse function $N(t)$. The absence of $H_{\alpha}$s due to the
vanishing of the $C^{\alpha}_{\beta\gamma}$s, implies that in
principle all $\gamma_{\alpha\beta}$s are candidates as arguments
for the wave function which solves the quadratic constraint
(Wheeler-DeWitt equation). This is in contrast to what happens in
other Bianchi Types \cite{Chris} where, less or equal to 3,
combinations of $\gamma_{\alpha\beta}$s and
$C^{\alpha}_{\beta\gamma}$s, parameterize the reduced
configuration space.

In this short communication, we present a complete reduction of
the initial configuration space for Bianchi Type I geometry --by
extracting as many gauge degrees of freedom, as possible. Two
separate cases are considered; when the cosmological constant is
present and when is not. In either case, a wave function which
depends on one degree of freedom (namely the determinant of the
scale factor matrix) is found, by imposing on it, the quantum
versions of all classical integrals of motion as additional
conditions.
\section{Bianchi Type I, the $\Lambda \neq 0$ case}
In this work, we will quantize the known action corresponding to
the most general Bianchi Type I cosmologies, i.e. the action
giving Einstein's Field Equations derived from the line element:
\be \label{lineelement}
ds^{2}=(-N^{2}(t)+N_{\alpha}(t)N^{\alpha}(t))dt^{2}+2N_{\alpha}(t)\sigma^{\alpha}_{i}(x)dx^{i}dt+
\gamma_{\alpha\beta}(t)\sigma^{\alpha}_{i}(x)\sigma^{\beta}_{j}(x)dx^{i}dx^{j}
\ee where $\sigma^{\alpha}_{i}$ are the invariant basis one-forms
of the homogeneous surfaces of simultaneity $\Sigma_{t}$,
satisfying: \be
d\sigma^{\alpha}=C^{\alpha}_{\beta\gamma}~\sigma^{\beta}\wedge
\sigma^{\gamma}\Leftrightarrow \sigma^{\alpha}_{i
,~j}-\sigma^{\alpha}_{j,~i}=2C^{\alpha}_{\beta\gamma}~
\sigma^{\gamma}_{i}~\sigma^{\beta}_{j} \ee with
$C^{\alpha}_{\beta\gamma}$ being the structure constants of the
corresponding isometry group. In 3 dimensions, the tensor
$C^{\alpha}_{\beta\gamma}$ admits a unique decomposition in terms
of a contravariant symmetric tensor density of weight $-1$,
$m^{\alpha\beta}$, and a covariant vector
$\nu_{\alpha}=\frac{1}{2}C^{\rho}_{\alpha\rho}$ as follows: \be
C^{\alpha}_{\beta\gamma}=m^{\alpha\delta}\varepsilon_{\delta\beta\gamma}+\nu_{\beta}\delta^{\alpha}_{\gamma}-\nu_{\gamma}\delta^{\alpha}_{\beta}.
\ee For the Class A $(\nu_{\alpha}=0)$ Bianchi Type I, this matrix
is \cite{Ellis}: \be m^{\alpha\beta}=\left(\begin{array}{ccc}
  0 & 0 & 0 \\
  0 & 0 & 0 \\
  0 & 0 & 0
\end{array}\right)
\ee resulting in vanishing structure constants.

As is well known \cite{Sneddon}, the Hamiltonian of the above
system is $H=\widetilde{N}(t)H_{0}+N^{\alpha}(t)H_{\alpha}$,
where: \be \label{hamiltonian}
H_{0}=\frac{1}{2}L_{\alpha\beta\mu\nu}\pi^{\alpha\beta}\pi^{\mu\nu}+\gamma
R +\gamma\Lambda\ee is the quadratic constraint, with: \be
\begin{array}{ll} \label{supermetricandR}
  L_{\alpha\beta\mu\nu}=\gamma_{\alpha\mu}\gamma_{\beta\nu}+\gamma_{\alpha\nu}\gamma_{\beta\mu}-
\gamma_{\alpha\beta}\gamma_{\mu\nu} \\
  R=C^{\beta}_{\lambda\mu}C^{\alpha}_{\theta\tau}\gamma_{\alpha\beta}\gamma^{\theta\lambda}
\gamma^{\tau\mu}+2C^{\alpha}_{\beta\delta}C^{\delta}_{\nu\alpha}\gamma^{\beta\nu}+
4C^{\mu}_{\mu\nu}C^{\beta}_{\beta\lambda}\gamma^{\nu\lambda}
\end{array}
\ee $\gamma$ being the determinant of $\gamma_{\alpha\beta}$, and:
\be \label{linearconstraint}
H_{\alpha}=C^{\mu}_{\alpha\rho}\gamma_{\beta\mu}\pi^{\beta\rho}
\ee are the linear constraints. Note that $\widetilde{N}$
appearing in the Hamiltonian, is to be identified with
$N/\sqrt{\gamma}$. For all Class A Types, the canonical equations
of motion, following from (\ref{hamiltonian}), are equivalent to
Einstein's equations derived from line element (\ref{lineelement})
--see \cite{Sneddon}.

The quantities $H_{0}$, $H_{\alpha}$ are weakly vanishing
\cite{Dirac}, i.e. $H_{0}\approx 0$, $H_{\alpha}\approx 0$. For
all Class A Bianchi Types ($C^{\alpha}_{\alpha\beta}=0$), it can
be seen --using the basic PBR $\{\gamma_{\alpha\beta},
\pi^{\mu\nu}\}=\delta^{\mu\nu}_{\alpha\beta}$-- that these
constraints are first class, obeying the following algebra \be
\label{algebra}
\begin{array}{l}
  \{H_{0}, H_{\alpha}\}=0 \\
  \{H_{\alpha},
  H_{\beta}\}=-\frac{1}{2}C^{\gamma}_{\alpha\beta}H_{\gamma},
\end{array}
\ee which ensures their preservation in time, i.e.
$\dot{H}_{0}\approx 0$, $\dot{H}_{\alpha}\approx 0$, and
establishes the consistency of the action.

If we follow Dirac's general proposal \cite{Dirac} for quantizing
this action, we have to turn $H_{0}$, $H_{\alpha}$, into operators
annihilating the wave function $\Psi$.

In the Schr\"{o}dinger representation: \be
\begin{array}{l}
  \gamma_{\alpha\beta}\rightarrow
\widehat{\gamma}_{\alpha\beta}=\gamma_{\alpha\beta} \\
  \pi^{\alpha\beta}\rightarrow
\widehat{\pi}^{\alpha\beta}=-i\frac{\partial}{\partial\gamma_{\alpha\beta}},
\end{array}
\ee with the relevant operators satisfying the basic Canonical
Commutation Relations (CCR) --corresponding to the classical ones:
\be [\widehat{\gamma}_{\alpha\beta},
\widehat{\pi}^{\mu\nu}]=i\delta^{\mu\nu}_{\alpha\beta}=\frac{i}{2}
(\delta^{\mu}_{\alpha}\delta^{\nu}_{\beta}+\delta^{\mu}_{\beta}\delta^{\nu}_{\alpha}).
\ee

In Bianchi Type I $(C^{\alpha}_{\beta\gamma}=0)$, the second of
(\ref{supermetricandR}) gives $R=0$, while relations
(\ref{linearconstraint}) vanish identically, and the algebra
(\ref{algebra}) is trivially satisfied. Thus, (\ref{hamiltonian}),
reads: \be \label{newhamiltonian}
H_{0}=\frac{1}{2}L_{\alpha\beta\mu\nu}\pi^{\alpha\beta}\pi^{\mu\nu}+\gamma\Lambda
\ee

Thus, the only operator which must annihilate the wave function,
is $\widehat{H}_{0}$; and the Wheeler-DeWitt equation
$\widehat{H}_{0}\Psi=0$, will produce a wave function, initially
residing on a 6-dimensional configuration space --spanned by
$\gamma_{\alpha\beta}$ s. The discussion however, does not end
here. If the linear constraints existed, a first reduction of the
initial configuration space, would take place \cite{Hajiceck}. New
variables, instead of the 6 scale factors, would emerge --say
$q^{i}$, with $i<6$. Then a new `'physical`' metric would be
induced: \be \label{physicalmetric}
g^{ij}=L_{\alpha\beta\mu\nu}\frac{\partial q^{i} }{\partial
\gamma_{\alpha\beta}}\frac{\partial q^{j}}{\partial
\gamma_{\mu\nu}} \ee According to Kucha\v{r}'s and Hajicek's
\cite{Hajiceck} prescription, the `'kinetic`' part of $H_{0}$
would have to be realized as the conformal Laplacian (in order for
the equation to respect the conformal covariance of the classical
action), based on the physical metric (\ref{physicalmetric}). In
the presence of conditional symmetries, further reduction can take
place, a new physical metric would then be defined similarly, and
the above mentioned prescription, would have to be used after the
final reduction \cite{Chris,Kuchar}.

The case of Bianchi Type I, is an extreme example in which all the
linear constraints, vanish identically; thus no initial physical
metric, exists --another peculiarity reflecting the high spatial
symmetry of the model  under consideration. In compensation, a lot
of integrals of motion exist ant the problem of reduction, finds
its solution through the notion of \emph{`'Conditional
Symmetries`'}. These linear in momenta integrals of motion, if
seen as vector field on the configuration space spanned by
$\gamma_{\alpha\beta}$s, induce --through their integral curves--
motions of the form $\widetilde{\gamma}_{\alpha\beta}=
\Lambda^{\mu}_{\alpha}\Lambda^{\nu}_{\beta}\gamma_{\mu\nu},
\Lambda \in GL(3,\Re)$ (see section 2 of \cite{diff}) which not
only are identical to the action of spatial diffeomorphisms, but
also describe the action of the automorphism group --since
$GL(3,\Re)$ is the Aut(G) for Type I \cite{Harvey}.

The generators of this automorphism group, are (in a collective
form and matrix notation) the following 9 --one for each
parameter: \be \label{Lambdas} \lambda^{\alpha}_{(I)\beta}=\left(
\begin{array}{ccc}
  a  & \beta & \delta \\
  \epsilon & \zeta & \eta \\
  \theta & \sigma & \rho
\end{array}\right),~~~ I \in [1,\ldots,9]
\ee with the defining property: \be
C^{\alpha}_{\mu\nu}\lambda^{\kappa}_{\alpha}=C^{\kappa}_{\mu\sigma}\lambda^{\sigma}_{\nu}+C^{\kappa}_{\sigma\nu}\lambda^{\sigma}_{\mu}.
\ee Exponentiating all these matrices, one obtains the outer
automorphism group of Type I, since there is not Inner
Automorphism subgroup (all structure constants vanish).

For full pure gravity, Kucha\v{r} \cite{Kuchar} has shown that
there are no other first-class functions, homogeneous and linear
in the momenta, except the linear constraints. If however, we
impose extra symmetries (e.g. the Bianchi Type I --here
considered), such quantities may emerge --as it will be shown. We
are therefore --according to Dirac \cite{Dirac}-- justified to
seek the generators of these extra symmetries; their
quantum-operator analogues will be imposed as additional
conditions on the wave function. The justification for such an
action, is obvious since these generators correspond to spatial
diffeomorphisms --which are the covariance of the theory. Thus,
these additional conditions are expected to lead us to the final
reduction, by revealing the true degrees of freedom. Such
quantities are, generally, called in the literature
\emph{`'Conditional Symmetries`'} \cite{Kuchar}.

From matrices (\ref{Lambdas}), we can construct the linear --in
momenta-- quantities: \be \label{epsilons}
E_{(I)}=\lambda^{\alpha}_{(I)\beta}\gamma_{\alpha\rho}\pi^{\rho\beta}
\ee In order to write analytically these quantities, the following
base is chosen: \be \label{basis}
\begin{array}{ccc}
 \lambda_{1}= \left(\begin{array}{ccc}
  0 & 1 & 0 \\
  0 & 0 & 0 \\
  0 & 0 & 0
\end{array}\right), & \lambda_{2}=\left(\begin{array}{ccc}
  0 & 0 & 1 \\
  0 & 0 & 0 \\
  0 & 0 & 0
\end{array}\right), & \lambda_{3}=\left(\begin{array}{ccc}
  0 & 0 & 0 \\
  0 & 0 & 1 \\
  0 & 0 & 0
\end{array}\right) \\
  \lambda_{4}=\left(\begin{array}{ccc}
  0 & 0 & 0 \\
  0 & 0 & 0 \\
  0 & 1 & 0
\end{array}\right), & \lambda_{5}=\left(\begin{array}{ccc}
  0 & 0 & 0 \\
  0 & 0 & 0 \\
  1 & 0 & 0
\end{array}\right), & \lambda_{6}=\left(\begin{array}{ccc}
  0 & 0 & 0 \\
  1 & 0 & 0 \\
  0 & 0 & 0
\end{array}\right) \\
  \lambda_{7}=\left(\begin{array}{ccc}
  1 & 0 & 0 \\
  0 & -1 & 0 \\
  0 & 0 & 0
\end{array}\right), & \lambda_{8}=\left(\begin{array}{ccc}
  0 & 0 & 0 \\
  0 & -1 & 0 \\
  0 & 0 & 1
\end{array}\right), & \lambda_{9}=\left(\begin{array}{ccc}
  1 & 0 & 0 \\
  0 & 1 & 0 \\
  0 & 0 & 1
\end{array}\right)
\end{array}
\ee

It is straightforward to calculate the Poisson Brackets between
$E_{(I)}$ and $H_{0}$: \be \label{commutatorEH} \{E_{(I)},
H_{0}\}=-\gamma\Lambda\lambda^{a}_{a} \ee But, it holds that: \be
\dot{E}_{(I)}=\{E_{(I)},H_{0}\}=-\gamma\Lambda\lambda^{a}_{a}\ee
--the last equality emerging by virtue of (\ref{commutatorEH}).
Thus: \be \label{integralsofmotion}
\dot{E}_{(I)}=\{E_{(I)},H_{0}\}=0 \Rightarrow
E_{(I)}=K_{(I)}=\textrm{constants},~~~ I \in [1,\ldots,8] \ee We
therefore conclude that, \underline{the first eight quantities}
$E_{(I)}$, are first-class, and thus integrals of motion. Out of
the eight quantities $E_{(I)}$, only five are functionally
independent (i.e. linearly independent, if we allow for the
coefficients of the linear combination, to be functions of the
$\gamma_{\alpha\beta}$ s); numerically, they are all independent.

The algebra of $E_{(I)}$ can be easily seen to be: \be
\label{algebraofepsilons}
\{E_{(I)},E_{(J)}\}=-\frac{1}{2}C^{M}_{IJ}E_{(M)},~~~I,J,M \in
[1,\ldots,9] \ee where: \be \label{algebraoflambda}
[\lambda_{(I)},\lambda_{(J)}]=C^{M}_{IJ}\lambda_{(M)},~~~I,J,M \in
[1,\ldots,9] \ee the square brackets denoting matrix
commutation.\\
The non vanishing structure constants of the algebra
(\ref{algebraoflambda}), are found to be: \be \label{liealgebra}
\begin{array}{lllll}
  C^{2}_{13}=1 & C^{4}_{15}=-1 & C^{7}_{16}=1 & C^{1}_{18}=-1 & C^{1}_{17}=-2 \\
  C^{1}_{24}=1 & C^{7}_{25}=1 & C^{8}_{25}=-1 & C^{3}_{26}=-1 & C^{2}_{27}=-1 \\
  C^{2}_{28}=1 & C^{8}_{34}=-1 & C^{6}_{35}=1 & C^{3}_{37}=1 & C^{3}_{38}=2 \\
  C^{5}_{46}=1 & C^{4}_{47}=-1 & C^{4}_{48}=-2 & C^{5}_{57}=1 & C^{5}_{58}=-1 \\
  C^{6}_{67}=2 & C^{6}_{68}=1
\end{array}
\ee

At this point, in order to achieve the desired reduction, we
propose that the quantities $E_{(I)}$ --with $I \in
[1,\ldots,8]$-- must be promoted to operational conditions acting
on the requested wave function $\Psi$ --since they are first class
quantities and thus integrals of motion (see
(\ref{integralsofmotion})). In the Schr\"{o}dinger representation:
\be \label{epsilonuponPsi}
\widehat{E}_{(I)}\Psi=-i\lambda^{\tau}_{(I)\alpha}\gamma_{\tau\beta}\frac{\partial
\Psi}{\partial \gamma_{\alpha\beta}}=K_{(I)}\Psi,~~~I \in
[1,\ldots,8] \ee In general, systems of equations of this type,
must satisfy consistency conditions decreed by the Frobenius
Theorem: \be
\begin{array}{ccc}
  \widehat{E}_{(J)}\Psi=K_{(J)}\Psi & \Rightarrow & \widehat{E}_{(I)}\widehat{E}_{(I)}\Psi=K_{(I)}K_{(J)}\Psi\\
  \widehat{E}_{(I)}\Psi=K_{(I)}\Psi & \Rightarrow & \widehat{E}_{(J)}\widehat{E}_{(I)}\Psi=K_{(J)}K_{(I)}\Psi
\end{array}
\ee Subtraction of these two and usage of
(\ref{algebraofepsilons}), results in: \be \label{selectionrule}
K^{M}_{IJ}\widehat{E}_{(M)}\Psi=0 \Rightarrow
C^{M}_{IJ}K_{(M)}=0\ee This relation constitutes a selection rule
for the numerical values of the integrals of motion. In view of
the Lie Algebra (\ref{liealgebra}), selection rule
(\ref{selectionrule}) imposes that all $K$s vanish, i.e.
$K_{1}=\ldots=K_{8}=0$. This fact restores the action of the
diffeomorphisms as covariances of the quantum theory, in the sense
that now, we have conditions of the form
$\widehat{E}_{(I)}\Psi=0$. Instead, if we also had $E_{(9)}$ (as
is the case $\Lambda=0$) then $K_{9}$ would remain arbitrary. With
this outcome, and using the method of characteristics
\cite{Carabedian}, the system of the five functionally independent
P.D.E. s (\ref{epsilonuponPsi}), can be integrated. The result is:
 \be \label{wavefunction}
\Psi=\Psi(\gamma) \ee i.e. an arbitrary (but well behaved)
function of $\gamma$ --the determinant of the scale factor matrix.

\emph{A note is pertinent here; from basic abstract algebra, is
well known that the basis of a linear vector space, is unique
--modulo linear mixtures. Thus, although the form of the system
(\ref{epsilonuponPsi}) is base dependent, its solution
(\ref{wavefunction}), is base independent.}

The next step, is to construct the Wheeler-DeWitt equation which
is to be solved by the wave function (\ref{wavefunction}). The
degree of freedom, is 1; the $q=\gamma$. According to Kucha\v{r}'s
proposal \cite{Hajiceck}, upon quantization, the kinetic part of
Hamiltonian (\ref{newhamiltonian}) is to be realized as the
conformal Beltrami operator -- based on the induced physical
metric --according to (\ref{physicalmetric}), with $q=\gamma$: \be
g^{11}= L_{\alpha\beta\mu\nu}\frac{\partial \gamma }{\partial
\gamma_{\alpha\beta}}\frac{\partial \gamma}{\partial
\gamma_{\mu\nu}}=L_{\alpha\beta\mu\nu}\gamma^{2}\gamma^{\alpha\beta}\gamma^{\mu\nu}
\stackrel{\textrm{first of~}
(\ref{supermetricandR})}{=}-3\gamma^{2}\ee In the Schr\"{o}dinger
representation: \be
\frac{1}{2}L_{\alpha\beta\mu\nu}\pi^{\alpha\beta}\pi^{\mu\nu}
\rightarrow -\frac{1}{2}\Box^{2}_{c} \ee where: \be \label{box}
\Box^{2}_{c}=\Box^{2}=\frac{1}{\sqrt{g_{11}}}~\partial_{\gamma}
\{\sqrt{g_{11}}~g^{11}~\partial_{\gamma}\} \ee is the
1--dimensional Laplacian based on $g_{11}$ ($g^{11}g_{11}=1$).
Note that in 1--dimension the conformal group is totally contained
in the G.C.T. group, in the sense that any conformal
transformation of the metric can not produce any change in the
--trivial-- geometry and is thus reachable by some G.C.T.
Therefore, no extra term in needed in (\ref{box}), as it can also
formally be seen by taking the limit $d=1,~R=0$ in the general
definition: \bdm
\Box^{2}_{c}\equiv\Box^{2}+\frac{(d-2)}{4(d-1)}R=\Box^{2} \edm
Thus: \be H_{0}\rightarrow
\widehat{H}_{0}=-\frac{1}{2}(-3\gamma^{2}\frac{\partial^{2}}{\partial\gamma}
-3\gamma\frac{\partial}{\partial\gamma})+\Lambda\gamma \ee So, the
Wheeler-DeWitt equation --by virtue of (\ref{wavefunction})--,
reads: \be
\widehat{H}_{0}\Psi=\gamma^{2}\Psi''+\gamma\Psi'+\frac{2}{3}\gamma\Lambda\Psi=0
\ee The general solution to this equation, is: \be
\Psi(\gamma)=c_{1}J_{0}(2\sqrt{\frac{2\gamma\Lambda}{3}})
+c_{2}Y_{0}(2\sqrt{\frac{2\gamma\Lambda}{3}}) \ee where $J_{n}$
and $Y_{n}$, are the Bessel Functions of the first and second kind
respectively --both with zero argument-- and $c_{1},c_{2}$,
arbitrary constants.

Some comments on this wave function. Indeed, at first sight, the
fact that $\Psi$ depends only on one argument and particularly on
$\gamma$, seems to point to some undesirable degeneracy regarding
anisotropy; classically $\gamma$ can be gauged to $e^{t}$ and thus
it seems as though the anisotropy parameter does not enter $\Psi$
at all. If, however, we reflect thoroughly, we will realize that
this objection rests strongly on a --not generally accepted--
mingling of the classical notion of anisotropy and the
interpretation of the wave function. Indeed if we adopt the
interpretation that the wave function $\Psi$ (along with a
suitable measure), is to give weight to all configurations
parameterized by $\gamma_{\alpha\beta}$, then the anisotropic
configuration will, in general, acquire different probabilities.
The degeneracy occurs only between two different anisotropic
configurations with the same determinant $\gamma$. In compensation
the scheme proposed here, avoids the gauge degrees of freedom as
much as possible. The final probabilistic interpretation must
await the selection of a proper measure, a theme which will be
discussed in section 4.
\section{Bianchi Type I, the $\Lambda=0$ case}
This section mimics the previous one, with minor changes. Indeed,
if the cosmological constant $\Lambda$ is zero, some
changes will take place.\\
The first concerns the obvious alteration to the Hamiltonian
(\ref{newhamiltonian}), which now reads: \be
\label{secondhamiltonian}
H_{0}=\frac{1}{2}L_{\alpha\beta\mu\nu}\pi^{\alpha\beta}\pi^{\mu\nu}
\ee This consequently, causes an alteration to the Poisson Bracket
(\ref{commutatorEH}), which takes the form: \be
\label{alteredcommutatorEH} \{E_{(I)},H_{0}\}=0,~~~I \in
[1,\ldots,9] \ee while, (\ref{algebraofepsilons}) still holds.
Thus in the present case, there are nine integrals of motion
--instead of eight. Also, the P.D.E. system (\ref{epsilonuponPsi})
consists of nine members (instead of eight), but now out of the
nine quantities $E_{(I)}$, only six are functionally independent;
the previous five, plus the $E_{(9)}$. Again, using the method of
characteristics \cite{Carabedian}, the system of the six
functionally independent P.D.E. s (\ref{epsilonuponPsi}), can be
integrated. The result is:
 \be \label{secondwavefunction}
\Psi=c_{1}\gamma^{iK_{9}/3} \ee where $\gamma$ is the determinant
of the scale factor and $K_{9}$, the remaining constant
--according to selection rule (\ref{selectionrule}).

The fact that this wave function does not depend on any
combination of $\gamma_{\alpha\beta}$s in an arbitrary manner
(i.e. $\Psi$ is not an arbitrary function of
$\gamma_{\alpha\beta}$s), might be taken as an indication that no
reduced Wheeler-DeWitt equation can be written. On the other hand,
this wave function does contain an arbitrary (essential) constant,
which ought to be fixed by the dynamics. The puzzle can be solved
by the following compromise; the initial configuration space,
should be the mini-superspace i.e. we should write the
Wheeler-DeWitt equation, based on the supermetric
$L^{\alpha\beta\mu\nu}$.

In the Schr\"{o}dinger representation: \be
\frac{1}{2}L_{\alpha\beta\mu\nu}\pi^{\alpha\beta}\pi^{\mu\nu}
\rightarrow -\frac{1}{2}\Box^{2}_{c} \ee  Thus using
(\ref{Ricciscalar}), (\ref{linearterm}), (\ref{finalBeltrami}) for
$d=3$ and $\mathcal{D}=6$, one may find respectively --see
appendix: \be R=\frac{15}{2} \ee \be
L_{\alpha\beta\mu\nu}\Gamma^{\alpha\beta\mu\nu}_{\kappa\lambda}=
-3\gamma_{\kappa\lambda} \ee and: \be \Box^{2}_{c}=
L_{\alpha\beta\mu\nu}\frac{\partial}{\partial
\gamma_{\alpha\beta}\gamma_{\mu\nu}}+3\gamma_{\kappa\lambda}\frac{\partial}{\partial
\gamma_{\kappa\lambda}}+\frac{3}{2} \ee Then Kucha\v{r}'s proposal
for (\ref{secondhamiltonian}) reads: \be H_{0}\rightarrow
\widehat{H}_{0}=-\frac{1}{2}\left(L_{\alpha\beta\mu\nu}\frac{\partial}{\partial
\gamma_{\alpha\beta}\gamma_{\mu\nu}}+3\gamma_{\kappa\lambda}\frac{\partial}{\partial
\gamma_{\kappa\lambda}}+\frac{3}{2}\right) \ee

Substitution of the wave function (\ref{secondwavefunction}) in
the Wheeler-DeWitt equation $\widehat{H}_{0}\Psi=0$, with
$\widehat{H}_{0}$ given by the previous relation, determines the
constant $K_{9}$. The outcome is: \be
\Psi=c_{1}\gamma^{\sqrt{2}/2}+c_{2}\gamma^{-\sqrt{2}/2} \ee The
constants $c_{1}, c_{2}$, remain arbitrary and may be fixed after
the selection of a proper measure via normalizability
requirements.
\section{Measure and Probabilistic Interpretation}
In general, the issue of the selection of a measure, is an open
question. For us, an important element for selecting it, is the
conformal covariance; the supermetric $L^{\alpha\beta\mu\nu}$ is
known only up to rescalings, because instead of $\widetilde{N}(t)$
one can take any $\overline{N}(t)=\widetilde{N}(t)e^{-2\omega}$
(with $\omega=\omega(\gamma_{\alpha\beta})$) and consequently
$\overline{L}^{\alpha\beta\mu\nu}(t)=L^{\alpha\beta\mu\nu}(t)e^{2\omega}$.
This property, is also inherited to the physical metric
(\ref{physicalmetric}) and is the reason for the Kucha\v{r}'s
recipe, adopted in this work.

It is therefore mandatory for the proposed measure $\mu$, to be
such that the probability density $\mu|\Psi|^{2}$, be invariant
under these scalings. Recalling that $\overline{\Psi}=\Psi
e^{(2-\mathcal{D})\omega/2}$, we conclude that $\mu$ must scale as
$\overline{\mu}=\mu e^{(\mathcal{D}-2)\omega}$.

It is not difficult to imagine such a quantity: any product of
$E_{(I)\alpha\beta}$ with $E_{(J)\mu\nu}$ (where
$E_{(I)\alpha\beta}=1/2(\lambda^{\kappa}_{a}\gamma_{\kappa\beta}+(\alpha\leftrightarrow
\beta))$ are the components of the vector field inferred from the
integrals of motion $E_{(I)}$) has the desired property. Indeed
the $E_{(I)}$ s do not scale at all, while the supermetric scales
as mentioned before. The group metric
$\Theta_{IJ}=C^{F}_{IS}C^{S}_{JF}$ can serve to close the group
indices of $E_{(I)\alpha\beta}$. So, we arrive at the quantity:
\be \label{xi}
\xi=L^{\alpha\beta\mu\nu}\Theta^{IJ}E_{(I)\alpha\beta}E_{(J)\mu\nu}
\ee (where $\Theta^{IJ}$ is the inverse of the group metric). The
quantity $\xi$ transforms as: $\overline{\xi}=\xi e^{2\omega}$.
Thus we only need to take $\mu=\xi^{(\mathcal{D}-2)/2}$ Using the
Lie algebra (\ref{liealgebra}), one obtains: \be \Theta_{IJ}\equiv
C^{F}_{IS}C^{S}_{JF}=\left(\begin{array}{ccccccccc}
  0 & 0 & 0 & 0 & 0 & 6 & 0 & 0 & 0 \\
  0 & 0 & 0 & 0 & 6 & 0 & 0 & 0 & 0 \\
  0 & 0 & 0 & 6 & 0 & 0 & 0 & 0 & 0 \\
  0 & 0 & 6 & 0 & 0 & 0 & 0 & 0 & 0 \\
  0 & 6 & 0 & 0 & 0 & 0 & 0 & 0 & 0 \\
  6 & 0 & 0 & 0 & 0 & 0 & 0 & 0 & 0 \\
  0 & 0 & 0 & 0 & 0 & 0 & 6 & 12 & 0 \\
  0 & 0 & 0 & 0 & 0 & 0 & 12 & 6 & 0 \\
  0 & 0 & 0 & 0 & 0 & 0 & 0 & 0 & 0
\end{array}\right) \ee
Thus, in order to have an inverse, we restrict ourselves to the
non trivial $8\times 8$ subspace, and have: \be
\Theta^{IJ}=\left(\begin{array}{cccccccc}
  0 & 0 & 0 & 0 & 0 & 1/6 & 0 & 0  \\
  0 & 0 & 0 & 0 & 1/6 & 0 & 0 & 0  \\
  0 & 0 & 0 & 1/6 & 0 & 0 & 0 & 0  \\
  0 & 0 & 1/6 & 0 & 0 & 0 & 0 & 0  \\
  0 & 1/6 & 0 & 0 & 0 & 0 & 0 & 0  \\
  1/6 & 0 & 0 & 0 & 0 & 0 & 0 & 0  \\
  0 & 0 & 0 & 0 & 0 & 0 & 1/9 & -1/18  \\
  0 & 0 & 0 & 0 & 0 & 0 & -1/18 & 1/9
\end{array}\right) \ee
That means that in the sum of the expression (\ref{xi}), we
exclude the vector $E_{9}$. After a straightforward calculation,
one finds that: \be \xi=\frac{5}{12} \ee The quantity
$\xi^{(\mathcal{D}-2)/2}$ in each of the two cases (i.e. with and
without the cosmological constant), defines the final expression
for the measure $\mu$.

A drawback of the current measure, is the loss of the hermiticity
of the Wheeler-DeWitt operator. This however is not such a blander
because, the eigenvalues of this operator, are zero and thus loss
of hermiticity does not result in having complex eigenvalues. On
the other hand the following welcomed features are obtained:
\begin{itemize}
\item In the case $\Lambda \neq 0$, the wave function which results in normalizability,
is:\\
$\Psi=c_{1}J_{0}(2\sqrt{\frac{2\gamma\Lambda}{3}})$, where $J_{0}$
is the Bessel function of the first kind with zero argument. Since
this function has a branch cut discontinuity in the complex z
plane running from $-\infty$ to 0, we are naturally led to
consider $\Lambda$ to be positive definite --since $\gamma$ is
positive definite. Also, from the behaviour of the probability
density, it is inferred that models with small $\Lambda$ are much
more probable. The limit $\Lambda \rightarrow 0$, gives a constant
wave function and therefore constant probability.
\item In the case $\Lambda=0$, the wave function turns out to be non
normalizable --if we use the measure above. This can be thought as
an indication against the favour of such a model over the previous
one. Of course, one could conceive of measures, e.g. gaussian
ones, which would make the probability normalizable. Further
investigation on this issue is irrelevant in view of the plethora
of these possibilities.
\end{itemize}
\section{Discussion}
The present work is the last in a series of publications in which
stress has been laid on the spirit and the usage of
\emph{`'Conditional Symmetries`'} within the programm of the
Quantum Cosmology approximation towards the quantization of
Gravity. It was shown how these can be used in order to achieve a
reduction of the initial configuration space --a problem closely
related to the symmetry of a model in general, and in particularly
to the high spatial symmetry of Bianchi Type I model. Indeed, when
these Conditional Symmetries are imposed on the requested wave
function, i.e. the P.D.E. systems (\ref{epsilonuponPsi}), along
with their consistency conditions, a significant reduction of the
configuration space takes place, by extracting as much gauge
degrees of freedom, as possible. This is a sine qua non procedure
in view of the fact that these conditional symmetries in Quantum
Cosmology, are nothing but a part of the G.C.T. group as is
explicitly shown in \cite{diff}. At the same time, most of the
classical constants of motion, are set equal to zero --a fact not
encountered before.\\
The reduced space, is one dimensional and the degree of freedom
left is still gauge ($\gamma$ is a density) if seen from the
3-slice point of view. In both of the cases under consideration, a
wave function is gained and a kind of "proposed r\^{o}le" for it,
is exhibited. Only in the first case, the wave function is
normalizable. Also, in this case, the results are in agreement
with the latest estimations, concerning $\Lambda$, i.e. that it is
positive and relatively small. Another peculiarity is that in
$\Lambda\neq 0$ case, there is a non zero probability when $\gamma
\rightarrow 0$ but this it might be seen as a side-effect of the
"preferred" coordinate system. Indeed, when $\gamma \rightarrow
0$, the same limit is attained by the determinant of the
supermetric, as well. If however, one forms any metric invariant
in the minisuperspace, he will see that the initially detected
singularity, is apparent and is due to the choice of the
coordinate system of this space, only.\\
At this point a remark concerning the implicit functional form of
wave functions with respect to the 3-metric, is pertinent. In the
literature it is common to assume a wave function dependant upon a
diagonal metric. Such an assumption is certainly understandable
from the a classical point of view, but not justified quantum
mechanically for two reasons:
\begin{itemize}
\item[$R_{1}$] The wave function is supposed to give weight to all
states, not only to the classical ones. So, ab initio, it must
depend on all 6 scale factors $\gamma_{\alpha\beta}$.
\item[$R_{2}$] As it has been mentioned in \cite{diff,timeaut}, there is a
particular class of G.C.T.s, which preserve manifest homogeneity
of the line element of the generic Bianchi Homogeneous 3-space,
has a non trivial action on the configuration space spanned by
$\gamma_{\alpha\beta}$'s, and this action is that of the
Automorphism group. Its differential description, leads to the
vector fields (\ref{epsilons}). Classically, in some Bianchi Types
(e.g. Type $VIII$, $IX$) an off mass-shell diagonalization of the
scale factor matrix is possible, and in some others (e.g. Type
$I$) the diagonalization is possible only on mass-shell
\cite{timeaut}. If one passes to the quantum level, i.e. if one
uses the generators $\lambda^{\alpha}_{\rho}$ of the corresponding
simplifying automorphisms $\Lambda^{\alpha}_{\rho}$, to construct
vector fields of the type
$\lambda^{\alpha}_{\rho}\gamma_{\alpha\sigma}\frac{\partial}{\partial\gamma_{\rho\sigma}}$,
he will encounter a very different situation; imposing these
fields upon $\Psi$ depending on all $\gamma_{\alpha\beta}$s, he
\underline{will not} get a $\Psi$ depending on the $\{\gamma_{11},
\gamma_{22}, \gamma_{33}\}$ only. For example the automorphism
group for Bianchi Type $IX$ is $SO(3)$, so
$\Lambda^{\alpha}_{\rho}$ are the rotations, and their generators,
are: \bdm
\begin{array}{ccc}
 \lambda_{1}= \left(\begin{array}{ccc}
  0 & 1 & 0 \\
  -1 & 0 & 0 \\
  0 & 0 & 0
\end{array}\right), & \lambda_{2}=\left(\begin{array}{ccc}
  0 & 0 & -1 \\
  0 & 0 & 0 \\
  1 & 0 & 0
\end{array}\right), & \lambda_{3}=\left(\begin{array}{ccc}
  0 & 0 & 0 \\
  0 & 0 & 1 \\
  0 & -1 & 0
\end{array}\right)
\end{array}
\edm The corresponding vector fields (\ref{epsilons}) upon $\Psi$
do not imply that this $\Psi$, in principle, an arbitrary function
of $\{\gamma_{11}, \gamma_{22}, \gamma_{33}\}$ only --while at the
classical level, the arbitrary scale factor matrix is diagonalized
through the usage of three independent rotations.
\end{itemize}
On the issue of the measure, we have succeeded in constructing a
measure which respects the scalings of $H_{0}$ and gives a square
integrable wave function (in the $\Lambda\neq 0$ case), at the
cost of loosing hermiticity of the Wheeler-DeWitt operator. The
last is not a serious drawback of the whole scheme, because not
only the Wheeler-DeWitt equation, constitutes a zero eigenfunction
problem, but also, the eigenfunctions (i.e. the permissible wave
functions) are complex functions in general --the probability
density in defined as the measure times the norm of the wave
function.

Lastly we would like to point out that the usage of the above
techniques, is in contact with the work \cite{henneaux} where the
strong gravity limit is treated. The connection to the strong
gravity limit of all the other Bianchi Types, can be made much
more explicit. The key observation is that the automorphism groups
of all other Bianchi Types, are subsets of $GL(3,\Re)$ i.e. the
automorphism group of Bianchi Type I. Consequently, if we adopt
the point of view of imposing the appropriate subset of the linear
integrals (\ref{epsilons}) on the wave function, corresponding to
the automorphism group of the desired Bianchi Type, we will get a
wave function which will satisfy the strong gravity limit of the
Wheeler-DeWitt equation. For example, if we select the subset (in
collective form) with:
\begin{equation}\lambda^{\alpha}_{\beta}=\left(\begin{array}{ccc}
  \kappa+\mu & x & y \\
  0 & \kappa & \rho \\
  0 & \sigma & \mu
\end{array}\right)
\end{equation}
we will get a $\Psi$ depending on $\gamma$ and
$q=\gamma^{2}_{11}/2\gamma$ and thus arrive at the strong gravity
limit of equation (19) in the first of \cite{Chris}. Likewise for
all Bianchi Types.\\
This line of reasoning may also satisfy those which, driven by
anisotropy ``classical'' considerations, would like to have more
than one arguments in the wave function. Of course, our point of
view is that the less gauge freedom in the wave function the
better and hence we adopt the procedure all the operator analogues
of the integrals of motion (\ref{epsilons}).

\vspace{1.5cm} \textbf{Acknowledgements}\\
One of us (G. O. Papadopoulos) is currently a scholar of the Greek
State Scholarships Foundation (I.K.Y.) and acknowledges the
relevant financial support.
\newpage
\appendix
\section{Appendix}
In this appendix, we give some useful formulae, concerning the
mini-superspace.
\vspace{0.5cm}

Using the results of canonical analysis in a $(d+1)$--dimensional
manifold, endowed with the line element (\ref{lineelement}), one
arrives at the notion of mini-superspace spanned by
$\gamma_{\alpha\beta}$ s (co-ordinates), and having as covariant
metric the following: \be \label{supercovmetric}
L^{\alpha\beta\mu\nu}=\frac{1}{4}(\gamma^{\alpha\mu}\gamma^{\beta\nu}+\gamma^{\alpha\nu}\gamma^{\beta\mu}
-2\gamma^{\alpha\beta}\gamma^{\mu\nu}) \ee while the contravariant
metric, is defined as: \be \label{superconmetric}
L_{\alpha\beta\mu\nu}=(\gamma_{\alpha\mu}\gamma_{\beta\nu}+\gamma_{\alpha\nu}\gamma_{\beta\mu}
-\frac{2}{d-1}\gamma_{\alpha\beta}\gamma_{\mu\nu}) \ee in the
sense that: \be \label{supercovconmetric}
L^{\alpha\beta\kappa\lambda}L_{\kappa\lambda\mu\nu}=\delta^{\alpha\beta}_{\mu\nu}
\equiv
\frac{1}{2}(\delta^{\alpha}_{\mu}\delta^{\beta}_{\nu}+\delta^{\alpha}_{\nu}\delta^{\beta}
_{\mu}) \ee

The Christof\mbox{}fel symbols are defined as: \be
\label{Chistoffel}
\Gamma^{\alpha\beta\mu\nu}_{\kappa\lambda}=\frac{1}{2}L_{\kappa\lambda\rho\sigma}
\{L^{\rho\sigma\mu\nu,\alpha\beta}+L^{\alpha\beta\rho\sigma,\mu\nu}
-L^{\alpha\beta\mu\nu,\rho\sigma}\}\ee where: \bdm
L^{\alpha\beta\mu\nu,\rho\sigma}\equiv \frac{\partial
L^{\alpha\beta\mu\nu}}{\partial \gamma_{\rho\sigma}} \edm Combined
usage of (\ref{supercovmetric}), (\ref{superconmetric}),
(\ref{supercovconmetric}) and (\ref{Chistoffel}), gives: \be
\Gamma^{\alpha\beta\mu\nu}_{\kappa\lambda}=-\frac{1}{4}(
\gamma^{\alpha\mu}\delta^{\beta\nu}_{\kappa\lambda}+
\gamma^{\alpha\nu}\delta^{\beta\mu}_{\kappa\lambda}+
\gamma^{\beta\mu}\delta^{\alpha\nu}_{\kappa\lambda}+
\gamma^{\beta\nu}\delta^{\alpha\mu}_{\kappa\lambda}) \ee

In the same spirit, the Riemann tensor is defined as follows: \be
R^{\alpha\beta\rho\sigma\mu\nu}_{\kappa\lambda}=
\Gamma^{\alpha\beta\rho\sigma,\mu\nu}_{\kappa\lambda}-
\Gamma^{\alpha\beta\mu\nu,\rho\sigma}_{\kappa\lambda}+
\Gamma^{\alpha\beta\rho\sigma}_{\omega\xi}
\Gamma^{\omega\xi\mu\nu}_{\kappa\lambda}-
\Gamma^{\alpha\beta\mu\nu}_{\omega\xi}
\Gamma^{\omega\xi\rho\sigma}_{\kappa\lambda} \ee where: \bdm
\Gamma^{\alpha\beta\mu\nu,\rho\sigma}_{\kappa\lambda}\equiv
\frac{\partial
\Gamma^{\alpha\beta\mu\nu}_{\kappa\lambda}}{\partial
\gamma_{\rho\sigma}} \edm Contraction of $(\rho,\sigma)$ with
$(\kappa,\lambda)$ results in the Ricci tensor: \be
R^{\alpha\beta\mu\nu}=
\Gamma^{\alpha\beta\kappa\lambda,\mu\nu}_{\kappa\lambda}-
\Gamma^{\alpha\beta\mu\nu,\kappa\lambda}_{\kappa\lambda}+
\Gamma^{\alpha\beta\kappa\lambda}_{\omega\xi}
\Gamma^{\omega\xi\mu\nu}_{\kappa\lambda}-
\Gamma^{\alpha\beta\mu\nu}_{\omega\xi}
\Gamma^{\omega\xi\kappa\lambda}_{\kappa\lambda} \ee A lengthy but
straightforward calculation, gives: \be \label{superRicci}
R^{\alpha\beta\mu\nu}=\frac{1}{8}(d\gamma^{\alpha\mu}\gamma^{\beta\nu}
+d\gamma^{\alpha\nu}\gamma^{\beta\mu}-2\gamma^{\alpha\beta}\gamma^{\mu\nu})
\ee

With the help of (\ref{superRicci}) and (\ref{superconmetric}) the
Ricci scalar is found to be: \be \label{Ricciscalar}
R=L_{\alpha\beta\mu\nu}R^{\alpha\beta\mu\nu}=\frac{1}{4}(d^{3}+d^{2}-2d)
\ee

Finally, the conformal Beltrami opetaror, is: \be \label{Beltrami}
\Box^{2}_{c}\equiv
\Box^{2}+\frac{\mathcal{D}-2}{4(\mathcal{D}-1)}=L_{\alpha\beta\mu\nu}
\frac{\partial^{2}}{\partial
\gamma_{\alpha\beta}\gamma_{\mu\nu}}-L_{\alpha\beta\mu\nu}\Gamma^{\alpha\beta\mu\nu}_{\kappa\lambda}
\frac{\partial}{\partial
\gamma_{\kappa\lambda}}+\frac{\mathcal{D}-2}{4(\mathcal{D}-1)}R
\ee where $\mathcal{D}$ is the dimension of the general metric
space: $\mathcal{D}=\frac{d(d+1)}{2}$, i.e. the number of the
independent
$\gamma_{\mu\nu}$.\\
One can find that: \be \label{linearterm}
L_{\alpha\beta\mu\nu}\Gamma^{\alpha\beta\mu\nu}_{\kappa\lambda}=
\frac{3-d^{2}}{d-1}\gamma_{\kappa\lambda} \ee thus
(\ref{Beltrami}), takes the form: \be \label{finalBeltrami}
\Box^{2}_{c}\equiv L_{\alpha\beta\mu\nu}
\frac{\partial^{2}}{\partial
\gamma_{\alpha\beta}\gamma_{\mu\nu}}-\frac{3-d^{2}}{d-1}\gamma_{\kappa\lambda}
\frac{\partial}{\partial\gamma_{\kappa\lambda}}
+\frac{\mathcal{D}-2}{4(\mathcal{D}-1)}R \ee

\end{document}